\documentclass[english,prl,twocolumn,showpacs]{revtex4}
\usepackage[T1]{fontenc}
\usepackage[latin9]{inputenc}
\usepackage{amsmath}
\usepackage{amssymb}
\usepackage{graphicx}
\usepackage{esint}
\usepackage[dvipdfm,
            colorlinks,
            linkcolor=red,
            citecolor=blue
            ]{hyperref}

\newcommand{\Eq}[1]{Eq.~\eqref{#1}}
\newcommand{\eq}[1]{\eqref{#1}}

\newcommand{\beq}{\begin{equation}}
\newcommand{\eeq}{\end{equation}}
\newcommand{\nn}{\nonumber}

\newcommand{\ket}[1]{\left\lvert{#1}\right\rangle}
\newcommand{\vect}[1]{\mathbf{#1}}
\newcommand{\sign}{\mathrm{sgn}}
\newcommand{\re}{\mathrm{Re}}
\newcommand{\im}{\mathrm{Im}}

\makeatletter
\@ifundefined{textcolor}{}
{%
 \definecolor{BLACK}{gray}{0}
 \definecolor{WHITE}{gray}{1}
 \definecolor{RED}{rgb}{1,0,0}
 \definecolor{GREEN}{rgb}{0,1,0}
 \definecolor{BLUE}{rgb}{0,0,1}
 \definecolor{CYAN}{cmyk}{1,0,0,0}
 \definecolor{MAGENTA}{cmyk}{0,1,0,0}
 \definecolor{YELLOW}{cmyk}{0,0,1,0}
}

\makeatother

\usepackage{babel}
\begin{document}

\title{Manipulation of $p$-wave scattering of cold atoms in low dimensions
using the magnetic field vector}

\author{Shi-Guo Peng$^{1}$}

\author{Shina Tan$^{2,3}$}

\email{shina.tan@physics.gatech.edu}

\author{Kaijun Jiang$^{1,3}$}

\email{kjjiang@wipm.ac.cn}

\affiliation{$^{1}$State Key Laboratory of Magnetic Resonance and Atomic and
Molecular Physics, Wuhan Institute of Physics and Mathematics, Chinese
Academy of Sciences, Wuhan 430071, China }

\affiliation{$^{2}$School of Physics, Georgia Institute of Technology, Atlanta,
Georgia, 30332, USA}

\affiliation{$^{3}$Center for Cold Atom Physics, Chinese Academy of Sciences,
Wuhan 430071, China}

\date{\today}
\begin{abstract}
It is well-known that the magnetic Feshbach resonances of cold atoms
are sensitive to the \emph{magnitude} of the external magnetic field.
Much less attention has been paid to the \emph{direction} of such
a field. In this work we calculate the scattering properties of spin
polarized fermionic atoms in reduced dimensions, near a $p$-wave Feshbach
resonance. Because of spatial anisotropy of the $p$-wave interaction,
the scattering has nontrivial dependence on both the magnitude and
the \emph{direction} of the magnetic field. In addition, we identify
an inelastic scattering process which is impossible in the isotropic-interaction
model; the rate of this process depends considerably on the direction
of the magnetic field. Significantly, an EPR entangled pair of identical
fermions may be produced during this inelastic collision.
This work opens a new method to manipulate resonant cold atomic interactions.
\end{abstract}

\pacs{03.75.Ss,05.30.Fk,67.85.-d,34.50.-s}

\maketitle
\emph{Introduction}.---Unlike electrons in condensed matter and nucleons
inside a nucleus, ultracold atoms have tunable interactions thanks
to the magnetic Feshbach resonances \cite{Chin2010F}.
The direction of the magnetic field plays little role in isotropic $s$-wave interactions,
other than providing a quantization axis for the hyperfine states.
For $p$-wave interactions, however,
the resonance positions for different orbital magnetic quantum numbers may be different
due to the anisotropic magnetic dipole-dipole interaction \cite{Ticknor2004M,Gunter2005P}.
The physical implication of this
split for cold atomic scatterings in reduced dimensions has not been explored theoretically.
In previous theoretical work on these scatterings, such split was
not taken into account \cite{Granger2004T,Pricoupenko2008R}. While
this omission is reasonable for atoms with small magnetic dipoles
such as $^{6}$Li \cite{Zhang2004P}, we have to consider
the effect of the split for other atoms such as $^{40}$K which show
large splits \cite{Ticknor2004M,Gunter2005P}.

In reduced dimensions, new two-body scattering resonances known as
confinement-induced resonances (CIRs) \cite{Olshanii1998A} appear. Recently an impressive amount of work
was devoted to the $s$-wave CIRs \cite{Haller,Frohlich2011R,Peng2010C,Peng2011C}.

In this Letter we study the scattering of two spin-polarized fermionic atoms
near $p$-wave Feshbach resonances, confined in low dimensions.
We will assume that the oscillator length
in the confinement directions is much larger than the tiny Van der Waals length scale.
Because of the split of resonance positions for different orbital magnetic quantum numbers,
we find that \textbf{1)} one can tune the scattering properties continuously by changing the direction
of the magnetic field, \textbf{2)} a new inelastic scattering
process where one of the two atoms is excited in the confined direction
by $\hbar\omega$ is now possible,
where $\omega$ is the angular frequency for the confinement, and \textbf{3)}
in quasi-one-dimension (quasi-1D), by choosing the magnitude and direction of the field and the collision energy judiciously,
one can have $100\%$ probability for this inelastic process,
thus creating an EPR entangled pair.

\begin{figure}
\includegraphics[width=1\columnwidth]{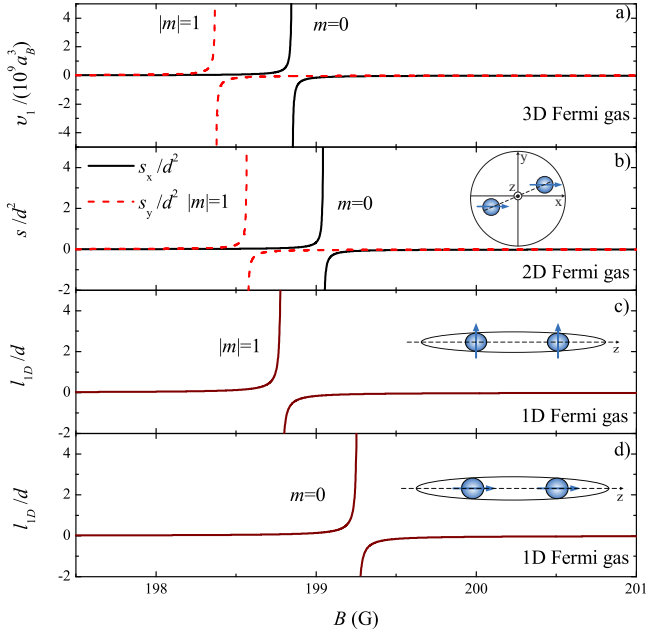}
\caption{(Color online) Resonant scattering properties of two spin-polarized
$^{40}$K fermionic atoms in the $\ket{F=9/2,m_F=-7/2}$ hyperfine state as functions of the external magnetic
field strength. (a) The $p$-wave scattering volume $v_{1m}$ in three dimensions \cite{Ticknor2004M}.
(b) The 2D scattering areas $s_x$ and $s_y$ in quasi-2D
at the scattering threshold ($\epsilon=0$), with the magnetic
field being parallel to the plane wherein atoms lie (i.e., $\alpha=\pi/2$).
(c) The 1D scattering length $l_{1D}$ in quasi-1D at the scattering threshold ($\epsilon=0$),
with the magnetic field being perpendicular to the waveguide (i.e.,
$\alpha=\pi/2$), and (d) the magnetic field is parallel to the waveguide
(i.e, $\alpha=0$). Here $a_{B}$ is the Bohr radius. 
To compare with the experiments,
we have chosen the corresponding parameters from Ref.~\cite{Gunter2005P},
and the effective-range-expansion parametrization from Ref.~\cite{Ticknor2004M}.}
\label{fig1}
\end{figure}

\textbf{Short-range boundary conditions}.---
In the presence of a uniform external magnetic field $\vect B$,
if two fermionic atoms in the same hyperfine state have a resonant $p$-wave interaction but negligible interactions in higher partial waves,
their relative wave function has a partial wave expansion
\beq
\psi(\vect r)=\sum_{m=-1}^{1}\Big[\frac{\mathcal{A}_{m}}{r^{2}}+\mathcal{B}_{m}+\mathcal{C}_{m}r+O(r^{2})\Big]Y_{1m}\left(\hat{\mathbf{r}}\right)
+O(r^3),\label{eq:S-1}
\eeq
at a distance $r$ that is small compared to the average inter-atomic distance but still large compared to the range of interaction.
Here $Y_{1m}(\hat{\vect r})$ are spherical harmonics whose north pole is in the direction of $\vect B$.

We assume that the scattering phase shifts $\delta_{1m}$ for different $m$'s may be different;
at small collision energies we shall take the following effective range expansion \cite{Ticknor2004M}:
\beq\label{effectiverange}
k^{3}\cot\delta_{1m}=-v_{1m}^{-1}+r_{1m}k^{2}/2,
\eeq
where $v_{1m}$ and $r_{1m}$ are respectively the $p$-wave scattering volume and effective range.
Unlike in the $s$-wave interaction, the $p$-wave effective range
is essential near resonance \cite{Idziaszek2006P,Peng2011,Madsen2002E}.
From \Eq{effectiverange} we can easily derive three linear constraints on the coefficients $\mathcal A_m$,
$\mathcal B_m$, and $\mathcal C_m$:
\begin{equation}
v_{1m}^{-1}\mathcal{A}_m-r_{1m}\mathcal{B}_{m}+3\mathcal{C}_{m}=0,\,\left(m=0,\pm1\right),\label{eq:S-2}
\end{equation}
which generalize the boundary condition in Ref.~\cite{Pricoupenko2006} to anisotropic $p$-wave interactions.
Unlike \Eq{effectiverange}, Eq.~\eq{eq:S-2} is also applicable to the quantum states without a certain collision energy.
In this sense \Eq{eq:S-2} is similar to the Bethe-Peierls boundary condition \cite{BethePeierls} for $s$-wave interactions.
We expect Eq.~\eq{eq:S-2} will be very useful in the theory of $N$ cold atoms with anisotropic $p$-wave interactions.

\textbf{Quasi-2D Fermi gases}.---
Consider the scattering of two fermionic atoms in the same hyperfine state, subject to a tight harmonic confinement along
the $z$ axis. The wave function $\psi(\vect r)$ for the relative motion
satisfies the Schr\"{o}dinger equation $(-\hbar^{2}\nabla^2/2\mu+\mu\omega^2z^2/2)\psi
=E\psi$ outside the range of interaction, where $\mu$ is the reduced mass, 
and $E$ is the relative collision energy which excludes
the center-of-mass energy. Without loss of generality we assume the
external magnetic field lies in the $xz$ plane, and forms an angle
$\alpha$ with the $+z$ axis. Assuming that the incoming atoms
are in the axial ground state, and the incident direction forms an angle
$\beta$ with the $+x$ axis, we obtain the scattering wave function
\begin{equation}
\psi(\mathbf{r})=\sin[q_{0}\rho\cos(\varphi-\beta)]e^{-\frac{z^{2}}{2d^{2}}}+\sum_{n}W_{n}r_{n}\mathcal{L}_{n}
(\epsilon,\mathbf{r}),\label{eq:2D-1}
\end{equation}
where $\epsilon\equiv\frac{ E}{\hbar\omega}-\frac{1}{2}$ for this quasi-2D geometry,
$\rho$ and $\varphi$ are polar coordinates: $\rho\cos\varphi=x$, $\rho\sin\varphi=y$,
$d=\sqrt{\hbar/\mu\omega}$ is the oscillator length, $q_0=\sqrt{2\epsilon}/d$,  the summation
is over $n=x,y,z$ with $r_x\equiv x$, $r_y\equiv y$, $r_z\equiv z$, and the
functions $\mathcal{L}_{n}(\epsilon,\vect {r}$) are related to the gradients of the Green's function
of the 3D harmonic oscillator Hamiltonian (see Appendix E of Ref.~\cite{Massignan2006T}):
\beq
\mathcal{L}_n(\epsilon,\vect r)=\frac{1}{\sqrt{2\pi}\,d^3}\int_0^\infty d\tau
\frac{e^{(\epsilon+\frac12)\tau-\frac{\rho^{2}}{2d^{2}\tau}-\frac{z^{2}}{2d^{2}\tanh\tau}}}
{\tau^{2-j_n}_{}\sinh^{j_n+\frac12}_{}\tau}
\eeq
for $\epsilon<j_n$.
Here $j_x=j_y=0$ and $j_z=1$.
To have outgoing scattered waves at higher energies, we should analytically continue $\mathcal{L}_{n}(\epsilon,\vect r)$
from $\epsilon<j_n$ to $\epsilon>j_n$ along routes \emph{above} the real-$\epsilon$ axis.

To determine the coefficients $W_n$ in \Eq{eq:2D-1}, we make a coordinate transformation
$x=x^{\prime}\cos\alpha+z^{\prime}\sin\alpha$, $y=y^{\prime}$,
$z=-x^{\prime}\sin\alpha+z^{\prime}\cos\alpha$. The $z'$ axis is parallel to the magnetic field.
Expanding $\psi$ at small $(x',y',z')$ and matching the boundary conditions \Eq{eq:S-2}, we find
\begin{subequations}\label{An}
\begin{align}
W_x&=-\frac{3d^3q_0(D_{0z}\cos^2\alpha+D_{1z}\sin^2\alpha)\cos\beta}{D_{0z}D_{1x}\cos^2\alpha+D_{0x}D_{1z}\sin^2\alpha},\\
W_y&=-\frac{3d^3q_0\sin\beta}{D_{1x}},\\
W_z&=\frac{3d^3q_0(D_0-D_1)\cos\beta\sin\alpha\cos\alpha}{D_{0z}D_{1x}\cos^2\alpha+D_{0x}D_{1z}\sin^2\alpha},
\end{align}
\end{subequations}
where
$D_{mn}= D_m+c_n(\epsilon)$,
$D_m= v_{1m}^{-1}d^3-(\epsilon+\tfrac12)r_{1m}d$, and
$c_n(\epsilon)=\lim_{r\to0}3d^3\big[\mathcal{L}_n(\epsilon,\vect r)-r^{-3}-(\epsilon+\tfrac12)/d^2r\big]$.
Note that $c_{n}(\epsilon)$ are pure mathematical functions \cite{cn2D}.

If $0<\epsilon<1$, both atoms will remain in the axial ground state after the collision,
i.e. the collision is purely elastic. At large $\rho$ the wave function takes
the form
\beq
\psi(\vect r)\approx-\frac{i}{2}e^{-\frac{z^2}{2d^2}}\Big\{
2i\sin\big[q_0\rho\cos(\varphi-\beta)\big]
+\frac{f_{00}(\varphi)}{\sqrt\rho}e^{iq_0\rho}\Big\},
\eeq
where $f_{00}(\varphi)$ is the elastic scattering amplitude:
\beq\label{f002D}
f_{00}(\varphi)=e^{\pi i/4}\big(2\sqrt{2q_0}\,/d\big)(W_x\cos\varphi+W_y\sin\varphi).
\eeq
Doing a partial-wave expansion at large $\rho$, we find $\psi$
is a linear combination of $e^{-z^2/2d^2}\rho^{-1/2}\cos(q_0\rho+\frac\pi4+\delta_x)\cos\varphi$
and $e^{-z^2/2d^2}\rho^{-1/2}\cos(q_0\rho+\frac\pi4+\delta_y)\sin\varphi$ plus higher partial wave components.
The phase shifts $\delta_x$ and $\delta_y$ are independent of the incident angle $\beta$:
\beq
e^{2i\delta_x}=1+\frac{i\sqrt\pi\,q_0}{d}\frac{W_x}{\cos\beta};~~~
e^{2i\delta_y}=1+\frac{i\sqrt\pi\,q_0}{d}\frac{W_y}{\sin\beta}.
\eeq
We find the energy-dependent \emph{scattering areas} for the $p$-wave scattering in two dimensions:
\begin{subequations}\begin{align}
s_x&=-\frac{\tan\delta_x}{q_0^2}=\frac{3\sqrt\pi\,d^2(D_{0z}\cos^2\alpha+D_{1z}\sin^2\alpha)}
{2\,\re(D_{0z}D_{1x}\cos^2\alpha+D_{0x}D_{1z}\sin^2\alpha)},\\
s_y&=-\frac{\tan\delta_y}{q_0^2}=\frac{3\sqrt\pi\,d^2}{2\,\re D_{1x}},
\end{align}\end{subequations}
where $\re$ stands for the real part.
The total elastic scattering cross section $\sigma_{00}=\frac12\int_0^{2\pi}|f_{00}(\varphi)|^2d\varphi$ yields
\beq\label{sigma002D}
\sigma_{00}=\frac{16}{q_{0}}\left(\sin^{2}\delta_{x}\cos^{2}\beta+\sin^{2}\delta_{y}\sin^{2}\beta\right).
\eeq

If $1<\epsilon<2$, one atom may go to the axial first excited state
after the collision. At large $\rho$ we have approximately
\begin{align}
\psi(\vect r)&\propto\big\{
2i\sin\big[q_0\rho\cos(\varphi-\beta)\big]
+f_{00}(\varphi)\rho^{-1/2}e^{iq_0\rho}\big\}\phi_0(z)\nn\\
&\quad+f_{01}\rho^{-1/2}e^{iq_1\rho}\phi_1(z),
\end{align}
where $q_1={\sqrt{2(\epsilon-1)}}\,/{d}$,
$\phi_0(z)=\frac{1}{\pi^{1/4}d^{1/2}}e^{-z^2/2d^2}$ and $\phi_1(z)=\frac{\sqrt{2}}{\pi^{1/4}d^{3/2}}ze^{-z^2/2d^2}$
are normalized axial energy eigenfunctions, $f_{00}(\varphi)$ is still given by \Eq{f002D},
and
$
f_{01}=4e^{3\pi i/4}W_z/(d^2\sqrt{q_1}\,).
$
The total scattering cross section is now
$\sigma=\sigma_{00}+\sigma_{01}$, with $\sigma_{00}$ being the elastic cross section
[still given by \Eq{sigma002D}], and $\sigma_{01}=\frac{q_1}{2q_0}\int_0^{2\pi}|f_{01}|^2d\varphi$
the inelastic cross section.
We find
\beq
\sigma_{01}=\frac{144\pi\sqrt{2\epsilon}\,d(D_0-D_1)^2\cos^2\beta\sin^2\alpha\cos^2\alpha}
{\big|D_{0z}D_{1x}\cos^2\alpha+D_{0x}D_{1z}\sin^2\alpha\big|^2}.
\eeq
The occurrence of this inelastic process has three necessary conditions:
(A) the phase shifts for $m=0$ and $|m|=1$ should be different, so that $D_0\ne D_1$;
(B) the magnetic field is neither parallel nor perpendicular to the quasi-2D plane, i.e., $\alpha\ne0,\pi/2$;
(C) the magnetic field is not perpendicular to the incident direction, so that $\beta\ne\pi/2$.

As a consistency check, one can verify that the total cross section obeys the optical theorem
in 2D \cite{OpticalTheorem2D}:
$\sigma=2\sqrt{2\pi/q_0}\,\mathrm{Re}\big[e^{-3\pi i/4}f_{00}(\beta)\big]$.

\begin{figure}
\includegraphics[width=1\columnwidth]{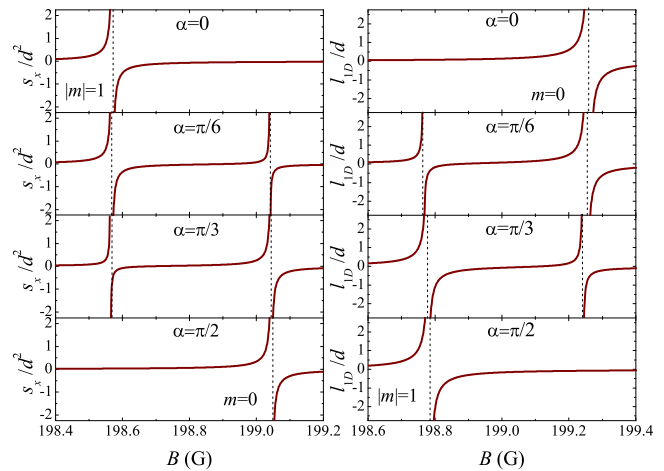}
\caption{\label{sl}
Left column: the 2D scattering area $s_x$ at $\epsilon=0$ for various angles
between the field and the normal direction of the quasi-2D plane.
Right column: the 1D scattering length $l_{1D}$ at $\epsilon=0$ for various angles between the field
and the quasi-1D line. Dashed lines indicate the locations of resonances.
Parameters are the same as in Fig.~\ref{fig1}.
}
\end{figure}

\begin{figure}
\includegraphics[width=1\columnwidth]{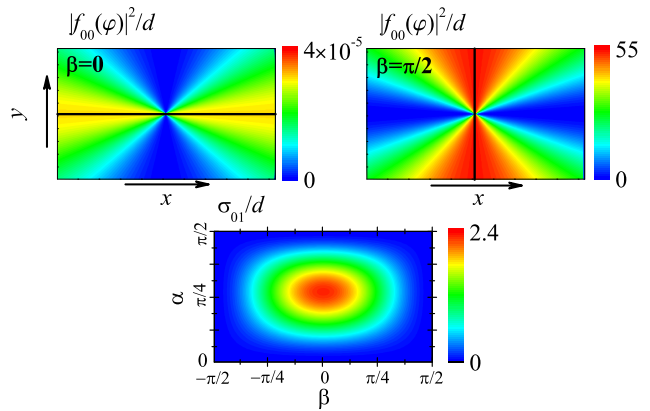}
\caption{(Color online) (Top) The angular distributions of the elastic differential
cross section $\left|f_{00}(\varphi)\right|^{2}$ 
in quasi-2D for different incident directions (black lines) at $\epsilon=0.01$, $\alpha=\pi/2$, and
$B=198.574\text{G}$. The corresponding scattering phase shifts are $\delta_x\approx -0.037^{\circ}$
and $\delta_y\approx- 59.588^{\circ}$. (Bottom) The inelastic
cross section $\sigma_{01}$ 
varying with the relative direction of the magnetic field at $\epsilon=1.5$ and $B=199.623$G.
All other parameters are the same as in Fig.~\ref{fig1}.}
\label{crosssection}
\end{figure}

Let us illustrate our results using the parameters in the quasi-2D $^{40}\mathrm{K}$ experiment \cite{Gunter2005P}.
In Fig.~\ref{fig1}b we plot the $p$-wave scattering areas versus the magnetic field strength at $\epsilon=0$.
The resonance positions are shifted from their free-space values, mainly due to the zero-point energy,
and they agree with Ref.~\cite{Gunter2005P}.
In Fig.~\ref{sl} (left column) we only show how the resonances in $s_x$ change as the magnetic field is tilted, since $s_y$ does not
depend on $\alpha$ ($s_y$ is shown in Fig.~\ref{fig1}b). If $\alpha=0$, there is only one resonance in $s_x$, corresponding to the $|m|=1$ channels. As $\alpha$ increases, one additional resonance in $s_x$ corresponding to the $m=0$ channel appears. The relative widths of the two resonances change, and their positions shift slightly as well. When $\alpha=\pi/2$, the $|m|=1$ resonance in $s_x$ disappears, while the $|m|=1$ resonance in $s_y$ still exists (see Fig.\ref{fig1}b).
In Fig.~\ref{crosssection} (top) we show the anisotropy of the $p$-wave interaction by
plotting the elastic differential cross section $|f_{00}(\varphi)|^2$ at $\epsilon=0.01$ and $B=198.574$G,
which is detuned by approximately $-0.001$G from the free space $|m|=1$ resonance at collision energy $0.51\hbar\omega$.
In Fig.~\ref{crosssection} (bottom) we plot the inelastic cross section $\sigma_{01}$
at $\epsilon=1.5$ and $B=199.623$G (detuned by approximately $-0.014$G from the free space $m=0$ resonance at collision energy $2\hbar\omega$).
We find that $\sigma_{01}$ has a broad peak
as a function of the \emph{direction} of the field relative to the incident direction
of the two atoms.

\textbf{Quasi-1D Fermi gases}.---We now consider the scattering of two fermionic atoms in the same hyperfine state,
subject to an axially symmetric harmonic confinement toward the $z$ axis.
The wave function $\psi_{1D}(\vect r)$ for the relative motion satisfies
the Schr\"{o}dinger equation $(-\hbar^2\nabla^2/2\mu+\mu\omega^2\rho^2/2)\psi_{1D}=E\psi_{1D}$
outside the range of interaction, where $\mu$ is the reduced mass, $\rho=\sqrt{x^2+y^2}$, and $E$
is the relative energy which excludes the center-of-mass energy.
Let the external magnetic field lie in the $xz$ plane, and form an angle $\alpha$ with the $+z$ axis.
Assuming that the incoming atoms are in the transverse ground state, we obtain the scattering wave function
\beq\label{eq:1D-1}
\psi_{1D}(\vect r)=e^{-\rho^2/2d^2}\sin(q_0z)+\sum_nW_n^{1D}r_n\mathcal{L}_n^{1D}(\epsilon,\vect r),
\eeq
where $\epsilon\equiv \frac{E}{\hbar\omega}-1$ for this quasi-1D geometry, $d=\sqrt{\hbar/\mu\omega}$, $q_0=\sqrt{2\epsilon}/d$,
the summation is over $n=x,y,z$, and
\beq
\mathcal{L}_n^{1D}(\epsilon,\vect r)=\frac{1}{\sqrt{2\pi}\,d^3}\int_0^\infty d\tau\frac{e^{(\epsilon+1)\tau-\frac{\rho^2}{2d^2\tanh\tau}-\frac{z^2}{2d^2\tau}}}
{\tau^{\frac32-j_n}_{}\sinh^{j_n+1}_{}\tau}
\eeq
for $\epsilon<j_n$.
Here $j_x=j_y=1$ and $j_z=0$.
To have outgoing scattered waves at higher energies, we analytically continue $\mathcal{L}_n^{1D}(\epsilon,\vect r)$
toward greater values of $\epsilon$ along routes \emph{above} the real-$\epsilon$ axis.
Taking the same coordinate transformation as in the quasi-2D case,
and using the boundary conditions \Eq{eq:S-2}, we find
\begin{subequations}\label{An1D}
\begin{align}
W_x^{1D}&=\frac{3d^3q_0(D_0^{1D}-D_1^{1D})\sin\alpha\cos\alpha}{D_{0z}^{1D}D_{1x}^{1D}\cos^2\alpha+D_{0x}^{1D}D_{1z}^{1D}\sin^2\alpha},\\
W_y^{1D}&=0,\\
W_z^{1D}&=-\frac{3d^3q_0(D_{1x}^{1D}\cos^2\alpha+D_{0x}^{1D}\sin^2\alpha)}
{D_{0z}^{1D}D_{1x}^{1D}\cos^2\alpha+D_{0x}^{1D}D_{1z}^{1D}\sin^2\alpha},
\end{align}
\end{subequations}
where
$D_{mn}^{1D}=D_m^{1D}+c_n^{1D}(\epsilon)$,
$D_m^{1D}=v_{1m}^{-1}d^3-(\epsilon+1)r_{1m}d$, and
$c_n^{1D}(\epsilon)=\lim_{r\to0}3d^3\big[\mathcal{L}_n^{1D}(\epsilon,\vect r)-r^{-3}-(\epsilon+1)/d^2r\big]$.
Note that $c_n^{1D}(\epsilon)$ are pure mathematical functions \cite{cn1D}.

For $0<\epsilon<2$, at $|z|\to\infty$ we find asymptotically
\begin{align}
\psi_{1D}(\vect r)&\propto\sign(z)\big(e^{-iq_0|z|}+g_{00}e^{iq_0|z|}\big)e^{-\rho^2/2d^2}\nn\\
&\quad+\theta(\epsilon-1)g_{01}e^{iq_1|z|}(\sqrt{2}\,x/d)e^{-\rho^2/2d^2},
\end{align}
where $\sign(z)$ is the sign of $z$, $\theta$ is the unit step function, $q_1=\sqrt{2(\epsilon-1)}/d$,
$g_{00}=-1-4iW_z^{1D}/d^2$, and
$g_{01}=4W_x^{1D}/\big(\sqrt{\epsilon-1}\,d^2\big)$.

In the window of energies $0<\epsilon<1$, the 1D scattering is purely elastic, with a 1D phase shift $\delta_{1D}$ given by the equation
$g_{00}=-e^{2i\delta_{1D}}$.
The energy-dependent 1D scattering length is 
\beq\label{l1D}
l_{1D}=-\frac{\tan\delta_{1D}}{q_0}=\frac{6d(D_{1x}^{1D}\cos^2\alpha+D_{0x}^{1D}\sin^2\alpha)}
{\bar{D}_{0z}^{1D}D_{1x}^{1D}\cos^2\alpha+D_{0x}^{1D}\bar{D}_{1z}^{1D}\sin^2\alpha},
\eeq
where $\bar D_{mz}^{1D}\equiv D_{mz}^{1D}-i6\sqrt{2\epsilon}=D_m^{1D}-12\zeta(-\frac12,\frac{2-\epsilon}{2})$,
and $\zeta(s,a)$ is the Hurwitz zeta function \cite{OlverHandbook}.

For isotropic interactions, $D_0^{1D}=D_1^{1D}$, our result reduces to $l_{1D}=6d/\bar D_{mz}^{1D}$,
which agrees with Ref.~\cite{Granger2004T} but disagrees somewhat with Ref.~\cite{Pricoupenko2008R}.
At $\epsilon=0$ we find $l_{1D}=6d/\big[(d/a_{1})^3-r_1d-12\zeta(-\frac12)\big]$, where the denominator contains
an extra term $-r_1d$ compared to Eq.(9) of Ref.~\cite{Pricoupenko2008R}, because of a finite zero-point energy $\hbar\omega$
for the collision.
For anisotropic interactions, $D_0^{1D}\ne D_1^{1D}$, we should use the more general formula \Eq{l1D},
and $l_{1D}$ depends on the direction of the magnetic field.
In Fig.~\ref{fig1} (c)(d) we illustrate this by plotting $l_{1D}$ for $^{40}$K at $\epsilon=0$ for two different directions of the field.
The atom loss peaks observed in Ref.~\cite{Gunter2005P} were about $0.2\sim0.3$G to the right of
the resonance positions shown in Fig.~\ref{fig1} (c)(d),
probably because of the finite Fermi energy and thermal effects in the experiment.
In Fig.~\ref{sl} (right column) we show how the resonances in $l_{1D}$ change as the magnetic field is tilted.
The physics is somewhat analogous to the 2D case, but we see more appreciable shifts of the resonance positions as $\alpha$ changes.

In the window of energies $1<\epsilon<2$, one atom may go to a transverse first excited state
after the collision. The transmission coefficient to this excited state is
\beq
\mathcal{T}=\frac{q_1}{q_0}|g_{01}|^2=\frac{288\sqrt{\frac{\epsilon}{\epsilon-1}}\,(D_0^{1D}-D_1^{1D})^2\sin^2\alpha\cos^2\alpha}
{\big|D_{0z}^{1D}D_{1x}^{1D}\cos^2\alpha+D_{0x}^{1D}D_{1z}^{1D}\sin^2\alpha\big|^2}.
\eeq
The probability of the elastic scattering is $\mathcal R=|g_{00}|^2$.
For $1<\epsilon<2$ one can verify that $\mathcal{R}+\mathcal{T}=1$.

\begin{figure}
\includegraphics[width=0.8\columnwidth]{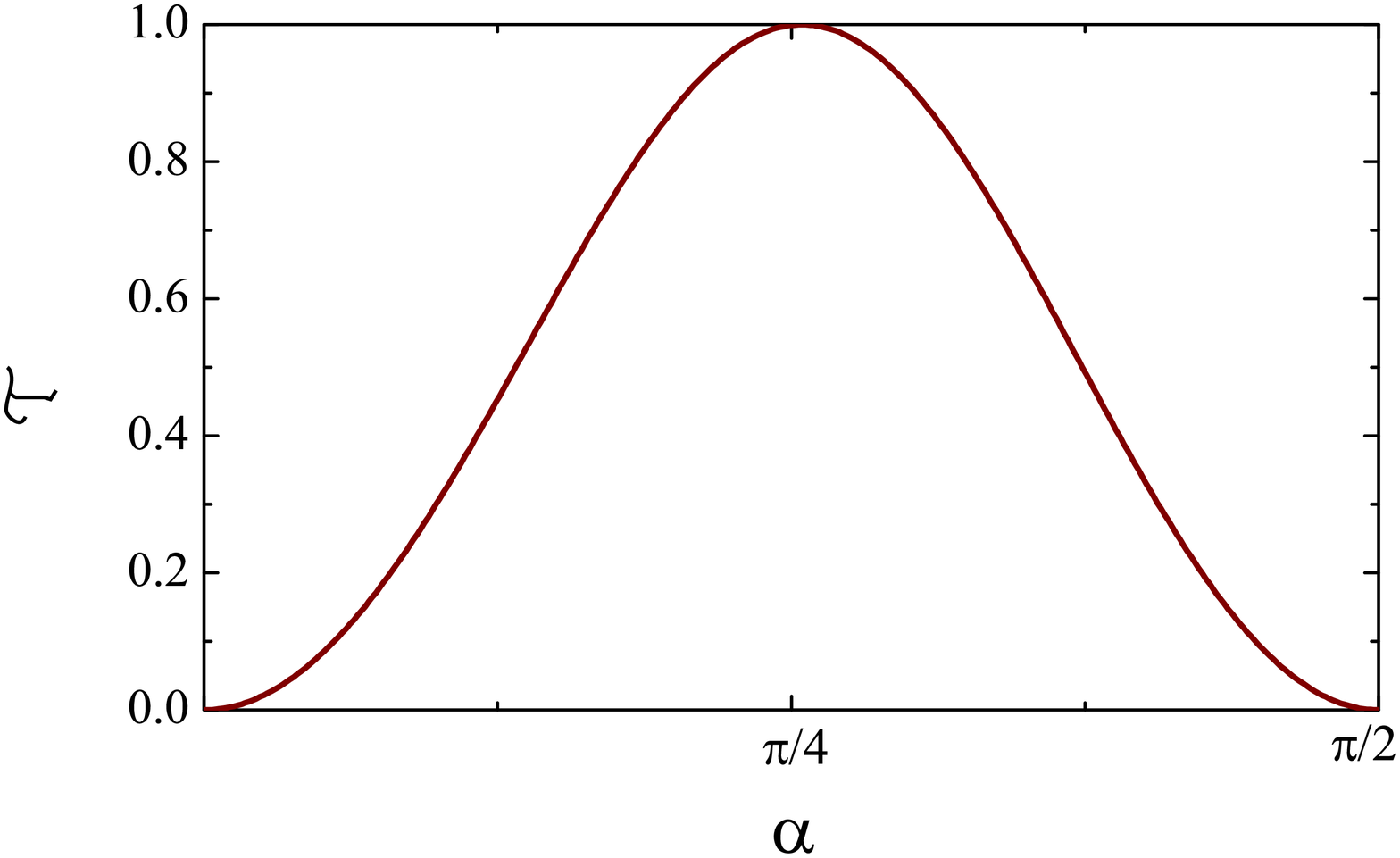}
\caption{The transmission coefficient to the transverse first
excited state for the collision of two fermionic atoms in quasi-1D at $\epsilon=1.5$ and $B=199.793\,\mathrm{G}$
(detuned by approximately $-0.039$G from the free space $m=0$ resonance at collision energy $2.5\hbar\omega$).
All other parameters are the same as in Fig.~\ref{fig1}.
The coefficient reaches unity at $\alpha\approx45.8^\circ$ in this plot.
}
\label{transmission}
\end{figure}

In Fig.~\ref{transmission} we plot the variation of the inelastic transmission coefficient with the direction of the magnetic field
at a \emph{magic value} of magnetic field strength, at which the peak value of $\mathcal{T}$ is 1.
At each collision energy satisfying $\epsilon_c<\epsilon<2$, we find two \emph{magic points} on the parameter plane $(B,\alpha)$,
such that
$
g_{00}=0
$
and the transmission coefficient $\mathcal T$ reaches unity  \cite{epsilon_c}. When this happens, one and only one atom
is excited transversely, but we do not know which, and the outgoing state
is an EPR pair with maximum entanglement, with two-body wave function
\beq
\Psi\!\propto\! e^{iq_1|z_1-z_2|}\!
\big[\phi_0(x_1,y_1)\phi_x(x_2,y_2)-\phi_x(x_1,y_1)\phi_0(x_2,y_2)\big],
\eeq
where $(x_i,y_i,z_i)$ are the coordinates of the $i$th atom, $\phi_0$ is the transverse ground state,
and $\phi_x(x,y)\propto x\phi_0(x,y)$ is a transverse excited state wave function.
The two magic points approach each other when $\epsilon$ approaches $\epsilon_c$ \cite{epsilon_c} from above,
merge into one point at $\epsilon=\epsilon_c$, and disappear at $\epsilon<\epsilon_c$.
When $\epsilon<\epsilon_c$ we always have $\mathcal{T}<1$.
When $\epsilon$ approaches 1 from above, $\mathcal{T}\propto(\epsilon-1)^{1/2}$, consistent with the threshold law for
inelastic collisions in 1D \cite{Sadeghpour2000Review}.

In conclusion, we have theoretically studied the low-dimensional scattering
properties of a spin polarized Fermi gas near $p$-wave Feshbach resonances.
Due to the spatial anisotropy of the $p$-wave interaction, the scattering
properties are strongly dependent on the direction of the external magnetic
field. We have also investigated the new inelastic scattering processes which
are absent in previous theoretical models based on isotropic $p$-wave interactions.
In particular, for the quasi-one-dimensional geometry, one can create an EPR entangled state
of two identical fermions using the collision; this may be useful for quantum information
with cold atoms.

\begin{acknowledgments}
S.-G. P. and K.-J. J. are supported by NSFC (No. 11004224, No. 11204355 and No. 91336106), NBRP-China (No. 2011CB921601), CPSF (No. 2012M510187 and No. 2013T60762) and programs in Hubei province (No. 2013010501010124 and No. 2013CFA056). S. T. is supported by the NSF (Grant No. PHY-1068511) and by the Alfred P. Sloan Foundation. We thank Shangguo Zhu for discussions.
\end{acknowledgments}


\begin{thebibliography}{10}
\bibitem{Chin2010F}C. Chin, R. Grimm, P. Julienne, and E. Tiesinga, Rev. Mod. Phys. \textbf{82},
1225 (2010).

\bibitem{Ticknor2004M}C. Ticknor, C. A. Regal, D. S. Jin, and J. L. Bohn, Phys. Rev. A \textbf{69},
042712 (2004).

\bibitem{Gunter2005P}K. G\"{u}nter, T. St\"{o}ferle, H. Moritz, M. K\"{o}hl, and T. Esslinger, Phys. Rev. Lett.
\textbf{95}, 230401 (2005).

\bibitem{Granger2004T}B. E. Granger and D. Blume, Phys. Rev. Lett.
\textbf{92}, 133202 (2004).

\bibitem{Pricoupenko2008R}L. Pricoupenko, Phys. Rev. Lett. \textbf{100},
170404 (2008).

\bibitem{Zhang2004P}J. Zhang, \emph{et al.}, Phys. Rev. A \textbf{70},
030702 (2004).

\bibitem{Olshanii1998A}M. Olshanii, Phys. Rev. Lett. \textbf{81},
938 (1998).

\bibitem{Haller}E. Haller, M. Gustavsson, M. J. Mark, J. G. Danzl, R. Hart, G. Pupillo, and H. C. N\"{a}gerl, Science \textbf{325}, 1224
(2009); E. Haller, M. J. Mark, R. Hart, J. G. Danzl, L. Reichs\"{o}llner, V. Melezhik, P. Schmelcher, and H.-C. N\"{a}gerl, Phys. Rev. Lett. \textbf{104}, 153203
(2010).

\bibitem{Frohlich2011R}B. Fr\"{o}hlich, M. Feld, E. Vogt, M. Koschorreck, W. Zwerger, and M. K\"{o}hl, Phys. Rev. Lett.
\textbf{106}, 105301 (2011).

\bibitem{Peng2010C}S.-G. Peng, S. S. Bohloul, X.-J. Liu, H. Hu, and P. D. Drummond, Phys. Rev. A \textbf{82},
063633 (2010); W. Zhang and P. Zhang, Phys. Rev. A \textbf{83}, 053615
(2011).

\bibitem{Peng2011C}S. G. Peng, H. Hu, X. J. Liu, and P. D. Drummond, Phys. Rev. A \textbf{84},
043619 (2011); S. Sala, P.-I. Schneider, and A. Saenz, Phys. Rev. Lett. \textbf{109},
073201 (2012); S. G. Peng, H. Hu, X. J. Liu, and K. J. Jiang, Phys. Rev. A \textbf{86},
033601 (2012).

\bibitem{Idziaszek2006P}Z. Idziaszek and T. Calarco, Phys. Rev. Lett.
\textbf{96}, 013201 (2006).

\bibitem{Peng2011}S. G. Peng, S. Q. Li, P. D. Drummond, and X. J. Liu, Phys. Rev. A \textbf{83},
063618 (2011); S. G. Peng, X. J. Liu, H. Hu, and S. Q. Li, Phys. Lett. A \textbf{375},
2979 (2011).

\bibitem{Madsen2002E}L. B. Madsen, Am. J. Phys. \textbf{70}, 811
(2002).

\bibitem{Pricoupenko2006}
L. Pricoupenko, Phys. Rev. Lett. \textbf{96}, 050401 (2006).

\bibitem{BethePeierls}
H. Bethe and R. Peierls,
Proc. R. Soc. A \textbf{148}, 146 (1935).

\bibitem{Massignan2006T}
P. Massignan and Y. Castin, Phys. Rev. A \textbf{74}, 013616 (2006).

\bibitem{cn2D}
One can evaluate $c_n(\epsilon)$ more easily by integrating the differential equations
$c_x''(\epsilon)=\frac{3\Gamma(-\frac\epsilon2)}{2\Gamma(\frac{1-\epsilon}{2})}$ and
$c_z'(\epsilon)=-\frac{6\Gamma(\frac{1-\epsilon}{2})}{\Gamma(-\frac{\epsilon}{2})}$
with initial conditions
$c_n(\epsilon)=(-2\epsilon-1)^{3/2}+O(|\epsilon|^{-1/2})$ at $\epsilon\to-\infty$.
We should analytically continue $c_n(\epsilon)$ toward greater values of $\epsilon$ by passing the isolated branch points
\emph{from above}. For instance, $\im\, c_x(\epsilon)=3\sqrt\pi\,\epsilon$ at $0<\epsilon<2$.


\bibitem{OpticalTheorem2D}
W. C. Henneberger, Phys. Rev. A \textbf{22}, 1383 (1980);
P. A. Maurone and T. K. Lim, Am. J. Phys. \textbf{51}, 856 (1983).

\bibitem{cn1D}
$c_x^{1D}(\epsilon)=3(\epsilon+1)\zeta\big(\tfrac12,\tfrac{1-\epsilon}{2}\big)+6\zeta\big(\!-\!\tfrac12,\tfrac{1-\epsilon}{2}\big)$ at $\epsilon<1$, and
$c_z^{1D}(\epsilon)=-12\zeta\big(\!-\!\tfrac12,-\tfrac{\epsilon}{2}\big)$ at $\epsilon<0$,
where $\zeta(s,a)$ is the Hurwitz zeta function \cite{OlverHandbook}.
We should analytically continue $c_n^{1D}(\epsilon)$ toward greater values of $\epsilon$ by passing the isolated branch points \emph{from above}.




\bibitem{OlverHandbook}
F. W. J. Olver, D. W. Lozier, R. F. Boisvert, C. W. Clark, NIST Handbook of Mathematical Functions,  Cambridge University Press, 2010.

\bibitem{epsilon_c}
For the trapping frequency in Ref.~\cite{Gunter2005P}, using the values of $v_{1m}$ and $r_{1m}$ from
Ref.~\cite{Ticknor2004M}, we find $\epsilon_c\approx1.00224$.

\bibitem{Sadeghpour2000Review}
H.~R.~Sadeghpour, J.~L.~Bohn, M.~J.~Cavagnero, B.~D.~Esry, I.~I.~Fabrikant, J.~H.~Macek, and A.~R.~P.~Rau,
J. Phys. B: At. Mol. Opt. Phys. \textbf{33}, R93 (2000).
\end{thebibliography}
\end{document}